\documentclass[12pt]{article} 
\usepackage{graphicx,subfigure,latexsym}
\usepackage{amsmath,amssymb,amsfonts}
\usepackage{bm}

\usepackage{color}

\usepackage{slashed} 
\usepackage{time}

\newcommand{\be}{\begin{equation}}
\newcommand{\ee}{\end{equation}}
\newcommand{\bear}{\begin{eqnarray}}
\newcommand{\eear}{\end{eqnarray}}
\newcommand{\ba}{\begin{array}}
\newcommand{\ea}{\end{array}}

\newcommand{\mq}{m_q}

\def \be {\begin{equation}}
\def \ee {\end{equation}}

\def \bes {\begin{subequations}}
\def \ees {\end{subequations}}

\def \pd {\partial}

\def \<{\langle}
\def \>{\rangle}
\def \+{\dagger}
\def \({\left(}
\def \){\right)}
\def \[{\left[}
\def \]{\right]}

\def \d {\delta}

\def \o {\omega}

\def \vq {\bm{q}}

\def \Re {\text{Re}}
\def \Im {\text{Im}}

\def \vac {\text{vac}}
\def \th {\text{th}}

\def \kapa {\kappa_{\parallel}}
\def \kape {\kappa_{\perp}}

\def \etapa {\eta_{\parallel}}
\def \etape {\eta_{\perp}}

\def \gluon {\,\textrm{gluon}}
\def \LLL {\,\textrm{LLL}}
\def \fermion {\,\textrm{fermion}}

\usepackage[normalem]{ulem}  

\renewcommand\sout{\bgroup \color{red} \ULdepth=-.5ex \ULset}
\newcommand{\blue}[1]{{\sf\color{blue}}}

\begin{document}

\begin{titlepage}

\begin{flushright}
{\normalsize RBRC-1157, RIKEN-QHP-210}\\
\end{flushright}




\begin{center}
{\Large\bf  Heavy Quark Diffusion in Strong Magnetic Fields
at Weak Coupling and Implications for Elliptic Flow}


\vskip 0.3in
Kenji Fukushima$^{1}$\footnote{e-mail: {\tt fuku@nt.phys.s.u-tokyo.ac.jp}},
Koichi Hattori$^{2,3}$\footnote{e-mail: {\tt koichi.hattori@riken.jp}},
Ho-Ung Yee$^{4,2}$\footnote{e-mail: {\tt hyee@uic.edu}}
and Yi Yin$^{5}$\footnote{e-email: {\tt yyin@bnl.gov}}
\vskip 0.15in

{\it $^{1}$ Department of Physics, The University of Tokyo, 7-3-1 Bunkyo-ku,
 Hongo, Tokyo 113-0033, Japan}\\[0.15in]
 {\it $^{2}$ RIKEN-BNL Research Center, Brookhaven National Laboratory, Upton,
 New York 11973-5000, U.S.A.}\\[0.15in]
 {\it $^{3}$ Nishina Center, RIKEN, Wako, Saitama 351-0198, Japan}\\[0.15in]
{\it $^{4}$ Department of Physics, University of Illinois, Chicago, Illinois
 60607, U.S.A.}\\[0.15in]
 {\it $^{5}$ Brookhaven National Laboratory, Upton,
 New York 11973-5000, U.S.A.}\\[0.15in]

 \today

\end{center}

\vfill

\begin{abstract}
We compute the momentum diffusion coefficients of heavy quarks, $\kappa_\parallel$ and
$\kappa_\perp$, in a strong magnetic field $B$ along the directions
parallel and perpendicular to $B$, respectively, at the leading order in QCD coupling constant
$\alpha_s$.  We consider a regime
relevant for the relativistic heavy ion collisions,
$\alpha_s eB\ll T^2\ll eB$, so that thermal excitations of light
quarks are restricted to the lowest Landau level (LLL) states.
In the vanishing light-quark mass limit, we find 
$\kappa_\perp^{\rm LO}\propto \alpha_s^2 T eB$ 
in the leading order that arises from screened Coulomb scatterings with (1+1)-dimensional LLL quarks,
while $\kappa_\parallel$ gets no contribution from the scatterings with LLL quarks due to kinematic restrictions. 
We show that the first non-zero leading order contributions to $\kappa_\parallel^{\rm LO}$ come from the two separate effects:  1) the screened Coulomb scatterings with
thermal gluons, and 2) a finite
light-quark mass $\mq$.  The former leads to
$\kappa_\parallel^{\rm LO,\,gluon}\propto \alpha_s^2 T^3$ and the
latter to $\kappa_\parallel^{\rm LO,\,massive}\propto
\alpha_s (\alpha_s eB)^{1/2} \mq^2$.  Based on our results,
we propose a new scenario for the large value of heavy-quark elliptic 
flow observed in RHIC and LHC. 
Namely, when $\kappa_\perp\gg\kappa_\parallel$, 
an anisotropy in drag forces gives rise to a sizable amount of the heavy-quark elliptic
flow even if heavy quarks do not fully belong to an ellipsoidally expanding background fluid. 
\end{abstract}

\vfill

\end{titlepage}
\setcounter{footnote}{0}

\baselineskip 18pt \pagebreak
\renewcommand{\thepage}{\arabic{page}}
\pagebreak


%
%

\section{Introduction}

Heavy ion collisions create Quark-Gluon Plasma (QGP) in the
presence of strong electromagnetic fields produced by charged
constituents of colliding nuclei~\cite{Skokov:2009qp,Voronyuk:2011jd,Bzdak:2011yy,Deng:2012pc}.
When the collision is non-central with a finite impact parameter, 
spectator protons produce a net magnetic field whose initial
strength could be comparable to the pion mass scale $eB \ge m_\pi^2$.
Experimental consequences from those enormous magnetic fields have
attracted much attention of theoretical studies (see
Refs.~\cite{review_Kha,Liao:2014ava, Miransky:2015ava, review_Hua} for
recent reviews).  One particular example is the chiral magnetic
effect (CME)~\cite{CME} 
from the interplay of the magnetic field and the quantum anomaly that
has been predicted to induce charge separation.  Closely related to
CME, it has been argued that the chiral magnetic
wave~\cite{Kharzeev:2010gd,Newman:2005hd} would cause an electric quadrupole moment
leading to charge-dependent elliptic flow~\cite{Burnier:2011bf,Gorbar:2011ya}.  Meanwhile, as attempts to
seek a signature of such strong $B$ with/without local parity
violation are made, possible enhanced anisotropic production of photons and dileptons has
been investigated in 
literature~\cite{Andrianov:2012hq,Tuchin:2012mf,BKS,FM,Yee:2013qma,HIS,Muller:2013ila,Yin}.

Whether the magnetic field leaves observable effects in heavy ion
collisions depends on several key properties in the early stage of
QGP. One of the crucial but still open questions is the
thermalization of light quarks in QGP that can potentially induce a
sizable electric conductivity.  If the electric conductivity is large
enough, decaying magnetic field would lead to an induced current and
this current would elongate the lifetime of the magnetic field
consistently with Lenz's law~\cite{Tuchin:2013apa,Tuchin:2015oka, MS}.
In turn, the strong magnetic field may also affect QGP thermalization
processes via coupling to the light quarks; i.e., the quark production
rate should depend on external electromagnetic
fields~\cite{Tanji:2010eu,Tuchin:2013ie,Fukushima:2015tza}.  
With these open questions in mind, exploring and studying observables
that are sensitive to the existence of the magnetic field would be
important, paving the way toward calibrating the strength and lifetime of 
the magnetic field and understanding interesting properties of QGP with the magnetic field.

In this work, we consider one of such important probes, namely,
dynamics of heavy quarks.  So far, several studies on magnetic field effects
have been carried out for \textit{static} properties of open heavy
flavors~\cite{Machado1, Machado2, Gub} and of
quarkonia~\cite{MT, AS, Cho, Guo, Bon, Rou, Dudal,Sadofyev:2015hxa}
\footnote{In contrast, by ``dynamical'' properties we mean real-time processes in a hot and dense medium. 
}. 
The measurements of open heavy flavors and quarkonia in the Relativistic Heavy
Ion Collider (RHIC) and the Large Hadron Collider (LHC), however, have
indicated the importance of \textit{dynamical} properties in the real-time
evolution inside the created matter, that is, transport and thermalization
of heavy flavors in the QGP~\cite{Moore:2004tg,HFR, Bera, Sca, Aka}
(see also Ref.~\cite{review_heavy} for a recent review and references
therein).  This motivates us to study heavy quark dynamics in QGP in
the presence of a strong magnetic field and to compute transport
coefficients that control the drag force and the time scale of
thermalization of heavy quarks. See Ref.\cite{RS} for a recent computation of heavy quark drag force in AdS/CFT correspondence with a {\it weak}, linearized magnetic field.

As in the previous studies of heavy quark diffusion without a magnetic
field~\cite{Moore:2004tg,CaronHuot:2007gq}, we use weak coupling
perturbative QCD techniques to compute the heavy quark diffusion constant in leading order (LO) of strong
coupling constant $\alpha_s$.   The heavy quark
mass $M_Q$ is assumed to be much larger than the scale of the magnetic
field; $M_Q\gg\sqrt{eB}$, which is a good assumption for charm and
bottom quarks, so the heavy quark motions are not directly
affected by magnetic fields\footnote{More precisely, the thermal
  velocity of heavy quark is $|\bm v|\sim \sqrt{T/M_Q}$, and the
  Lorentz force is $(d \bm p/dt)_{\rm Lorentz}= e\bm v\times\bm B$
  which is suppressed as $\sim eB\sqrt{T/M_Q}$ for large $M_Q$.  On
  the other hand, we will see that the momentum kicks from thermal
  scattering with light quarks and gluons that we compute at LO in
  this work are not suppressed for large $M_Q$.}.

At finite temperature, there exist thermally populated light quarks and gluons
that can scatter with the heavy quark, which gives random momentum
kicks to diffuse the heavy quark momentum.  At LO in $\alpha_s$ these
scatterings are mediated by one-gluon exchange and the scatterers can
be regarded as quasi-particles in thermally equilibrated matter.  For
magnetic field $eB\sim T^2$, it would put the light quark dispersion relation
into the Landau quantized ones (i.e. Landau levels, which will be
abbreviated as LL below) in thermal equilibrium, which will affect
both the scattering quark spectrum and its screening effect on the
one-gluon mediation via the gluon self-energy from quark loops.
We will show that the gluon screening mass (that is, the Debye mass) from the Landau quantized quark
one-loop is $m_{D,B}^2\sim \alpha_s eB$, whereas the one from the gluon one-loop is
$\alpha_s T^2$ as usual which is suppressed compared to the former quark contribution. 
Therefore, we include the screening mass $ m_{D,B}^2$ 
in the Coulomb scattering diagram between the heavy quark and thermal scatterers, 
which is necessary to regularize the infrared regime in the soft gluon exchanges.

The thermal masses of scatterers, i.e., time-like gluons and lowest Landau level (LLL) quarks, have the same order as $ m_{D,B}^2$. 
However, since the typical momenta of scattering quarks and gluons are hard $\sim T$, 
we assume that the self-energy corrections to  these hard LLL quarks and gluons in the present leading-order calculation 
can be neglected, which specifies a hierarchy $\alpha_s eB\ll T^2$. 
Note also that, in this regime, one can neglect the self-energy corrections to the hard thermal particles 
which compose the internal lines of the gluon self-energy diagrams, 
so that the screening mass $ m_{D,B}^2$ mentioned above 
can be obtained from the simple one-loop calculation.


Although the description of the case of $eB\sim T^2$ should
involve all the LLs of thermally equilibrated light
quarks (for the calculation of the self-energy with all LLs; see
Refs.~\cite{Feng:2012dqa,Fukushima:2012kc,Hattori:2012je}), we instead
consider an extreme limit of strong magnetic field $eB\gg T^2$ so that
only the LLL states are thermally occupied.  This allows us to obtain
some analytic results that are helpful to unravel the key physics.
Thus, in this work, the regime of our interest is specifically
given as $\alpha_s eB\ll T^2\ll eB$, and the higher LL
occupations are exponentially suppressed by powers of $e^{-\sqrt{eB}/ T}$.  In realistic heavy ion collisions, these inequalities are approximately
satisfied.

Our main finding is that in the presence of a strong magnetic field, the
drag forces or the momentum diffusion coefficients become highly
anisotropic.  In particular, the ratio between the longitudinal
momentum diffusion coefficient $\kappa_{\parallel}$ and the transverse
one $\kappa_{\perp}$ becomes
\be
 \label{kappa-ratio}
 \frac{\kappa_{\parallel}}{\kappa_{\perp}}\sim \frac{T^{2}}{eB}\;\ll 1\, ,
\ee
in the regime we are working on.  We will discuss the phenomenological
implication of \eqref{kappa-ratio} to heavy flavor elliptic flow and
will propose a new scenario for the sizable elliptic flow of heavy quarks observed in experiments:  a strong magnetic field would enhance
the heavy flavor elliptic flow even without thermalization of heavy quarks with respect to the expanding plasma, which is in favor of
resolving the heavy-flavor puzzle triggered by the elliptic flow
measurement of the $D$ mesons.
 
This paper is organized as follows.  In Section~\ref{sec:random} we
introduce the basic formulation of how to compute the momentum diffusion coefficients,
$\kappa_\parallel$ and $\kappa_\perp$, describing the in-medium heavy
quark motion.  At LO we express those transport coefficients using
the Coulomb scattering rate in terms of the longitudinal gluon
spectral function.  We then perform our explicit calculations of the
spectral function in Section~\ref{sec:computation}.  We present the
results in the zero quark mass limit in Section~\ref{sec:massless} and
find $\kappa_\perp\sim \alpha_{s}^2eB T$, while $\kappa_\parallel$ does not get such a contribution due to kinematic constraints. To
find the first non-vanishing contribution to $\kappa_\parallel$, we
then proceed to the hard gluon scattering contribution and also the finite
mass corrections in Section~\ref{sec:subleading}.  
In Section~\ref{sec:FP}, we discuss the phenomenological implication of our
results to a non-thermal origin of the heavy quark elliptic flow induced by strong
magnetic field.
We conclude in Section~\ref{sec:discussions}.

\section{Random Forces and Diffusion Coefficients}
\label{sec:random}

As a preparation for our computations in the subsequent sections,
let us here summarize the basic formalism for the heavy quark transports.
Heavy quarks in a finite temperature plasma are subject to random kicks from thermally excited
light quarks and gluons, and their motions are described by Langevin
equations as follows 
\cite{Moore:2004tg}:
\be
\label{eq:Langevin}
\frac {d p_z}{ dt} 
 = -\eta_\parallel p_z + \xi_z \,,
 \qquad
 {d\bm p_\perp\over dt} =
   -\eta_\perp \, \bm p_\perp + \bm\xi_\perp \,.
\ee
Since the external magnetic field provides a preferred spatial
direction,  we have a set of two equations for the heavy quark motions,
parallel and perpendicular to the magnetic field that is oriented in
the $z$-direction.  In Eq.~\eqref{eq:Langevin}, $p_z$ and $\bm p_\perp$ denote the heavy quark momenta parallel and perpendicular to the magnetic field, respectively, and the
random forces, $\xi_z$ and $\bm \xi_\perp$, as well as
the drag coefficients, $\eta_\parallel$ and $\eta_\perp$, should be
defined separately for parallel and perpendicular directions to the magnetic
field.  The random forces are assumed to be white noises,
\be
 \langle \xi_{z}(t) \xi_z(t')\rangle=\kappa_\parallel \delta(t-t')\,,\qquad
 \langle \xi_{\perp}^i(t)\xi_\perp^j(t')\rangle=\kappa_\perp
  \delta^{ij}\delta(t-t') \quad (i,j=x,y)
\, ,
\ee
and these coefficients, $\kappa_\parallel$ and $\kappa_\perp$, are
related to the drag coefficients, $\eta_\parallel$ and $\eta_\perp$,
through the fluctuation-dissipation theorem as
\be
\label{eq:fdtheorem}
  \eta_{\parallel} = 2 M_{Q}T \kappa_{\parallel} \, , \qquad 
  \eta_{\perp}= 2 M_{Q}T \kappa_{\perp}
\, . 
\ee
We will compute anisotropic momentum diffusion coefficients, $\kappa_\parallel$ and
$\kappa_\perp$, in the presence of magnetic
field.

The above Langevin picture as well as the separation between
longitudinal and transverse dynamics can be justified for a
sufficiently large $M_Q\gg eB/T$.  To see this, we should note that
the typical thermal momentum of heavy quark is of the order of
$\sqrt{M_Q T}$, and its typical velocity is
$|\bm v| \sim \sqrt{T/M_Q}$.  We will see in the following sections that the typical momentum
transfer $q_\perp=|\bm q_\perp|$ from the LLL quarks to the heavy quark in the LO computation ranges\footnote{
This latter inequality can be expected immediately from the
  size of the LLL wavefunction $\sim 1/\sqrt{eB}$ that is the inverse of
  the typical transverse momentum.} 
in $\sqrt{\alpha_s eB}\lesssim q_\perp \lesssim \sqrt{eB}$ 
and that the typical momentum transfer from thermal gluons at LO
ranges in $\sqrt{\alpha_s eB}\lesssim |\bm q| \lesssim T \ll \sqrt{eB}$.
Therefore, for $M_Q\gg eB/T\gg T$, the momentum transfer in each
scattering is small compared to the thermal momentum, and it involves many scatterings to
change the heavy quark momentum significantly.  Then, the description
of heavy quark motion should become statistical, leading to the above Langevin dynamics.  The same conclusion
can be obtained also by the condition that the energy transfer
$\omega$ in each collision should be much smaller than the
temperature, in order for the fluctuation-dissipation relation from
the equi-partition theorem to be meaningful, that is, 
$\omega\sim |\bm v\cdot \bm q|\sim\sqrt{T/M_Q}\cdot \sqrt{eB}\ll T$
holds when $M_Q\gg eB/T$.  The separation between the transverse and
the longitudinal dynamics in Eq.~\eqref{eq:Langevin} simply follows from knowing that the mixed
coefficient $\kappa_{\perp z}$ should be proportional to the transverse velocity $\bm v_\perp$ from rotational symmetry, which are of higher order in small velocity limit $|\bm v| \sim \sqrt{T/M_Q}\ll 1$.
\begin{figure}
     \begin{center}
              \includegraphics[width=0.7\hsize]{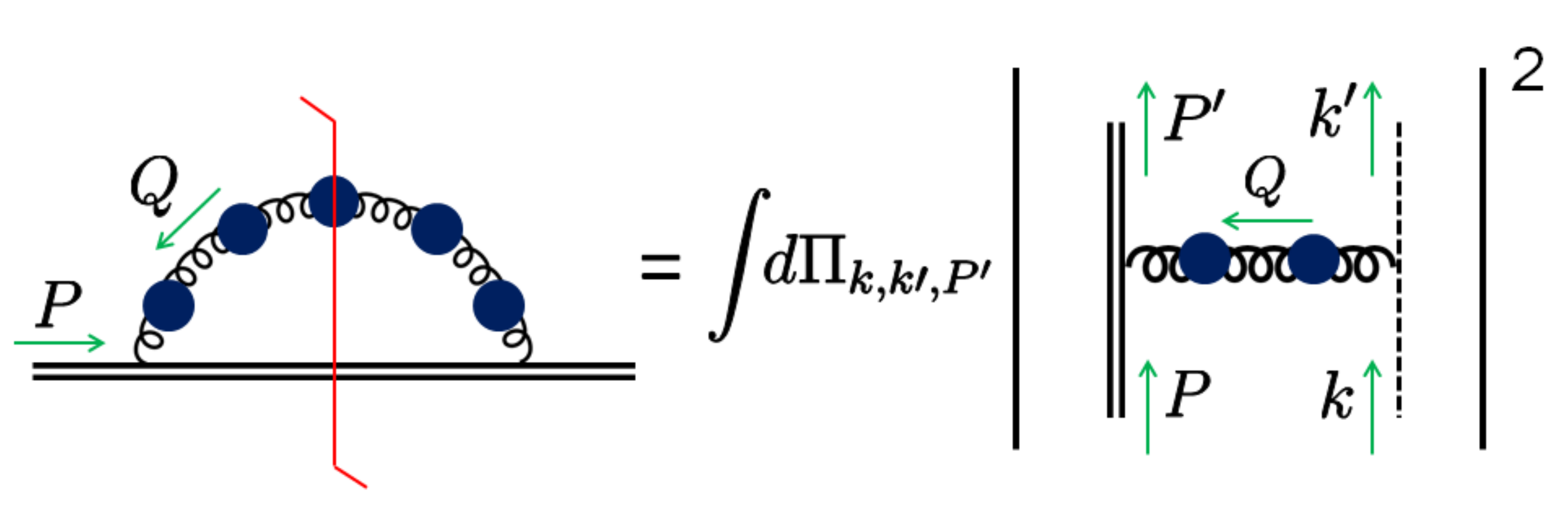}
     \end{center}
\vspace{-0.5cm}
\caption{Momentum diffusion of a heavy quark (double line) due to
  Coulomb scatterings with thermal quarks and gluons (collectively
  denoted as a dashed line).}
\label{fig:H-L}
\end{figure}

The momentum diffusion coefficients are equivalently defined by \be
\kappa_{ij}\equiv \lim_{\Delta t\to 0}{1\over 
\Delta t}{\langle \Delta p_i \Delta p_j\rangle}\,,\ee where $\Delta p^i=p^i(t+\Delta t)-p^i(t)$,
and these are interesting transport coefficients of the QGP medium.
They can be defined in a gauge invariant and non-perturbative way  as~\cite{CasalderreySolana:2006rq}
\begin{equation}
 \kappa_{ij} = \lim_{\omega\to0}\frac{4\pi\alpha_s}{d_H} \int_{-\infty}^{+\infty} dt\,
  e^{i\omega t}\,{\rm tr}\bigl\langle
  W(0,t) E_i(t) W(t,0) E_j(0)\bigr\rangle\,,
\end{equation}
where $E_i$ and $W$ are the chromoelectric field and the Wilson line
in the heavy quark color representation, respectively.  At LO in $\alpha_s$
the Wilson line is trivial and we can replace $E_i$ with
$\partial^i A^0$ in the static limit of $\omega\to0$.  The dimension
of the heavy quark representation $d_H$ is canceled by taking the
trace in color space, resulting in Casimir $C_R^{\rm HQ}$ given by
$C_R^{\rm HQ}=(N_c^2-1)/(2N_c)$ and $N_c$, respectively, for heavy
quarks in the fundamental and the adjoint representations.  Thus, we
need to evaluate the color diagonal part of the Wightman function of $A^0$ field, which is
denoted as $G^{>00}$. In
momentum representation spatial derivatives translate into
momenta, leading to 
\be
 \kappa_{ij} = \lim_{\omega\to0}{4\pi\alpha_s}C_R^{\rm HQ}\int{d^3\bm q\over (2\pi)^3} G^{>00}(\omega,\bm q)q_i q_j\,,
 \ee
 and from rotational symmetry, we have
 \be
 \kappa_{\parallel} = \int {d^3\bm q}\,
  {d\Gamma(\bm q)\over d^3\bm q}\, q_{z}^2 \,,
 \qquad
 \kappa_{\perp} = {1\over 2}\int {d^3\bm q}\,
  {d\Gamma(\bm q)\over d^3\bm q}\, \bm q_{\perp}^2 \,,
\label{kappa}
\ee 
where
\be
{d\Gamma(\bm q)\over d^3\bm q}\equiv{4\pi\alpha_s\over (2\pi)^3}C_R^{\rm HQ}\lim_{\omega\to0}G^{>00}(\omega,\bm q)=
{4\pi\alpha_s \over (2\pi)^3}C_R^{\rm HQ} \lim_{\omega\to0}{T\over\omega}\rho_L(\omega,\bm q)\,,\label{scatt}
\ee
can be interpreted as the scattering rate of the heavy quark via one-gluon exchange with thermal particles per
unit volume of momentum transfer $\bm q$. The $\rho_L$ is the longitudinal gluon spectral density and in the last equality we used a thermal relation
$G^{>00}(\omega)=n_B(\omega)\rho_L(\omega)$ which can be
expanded as $G^{>00}(\omega)\approx(T/\omega)\rho_L(\omega)$ for $\omega\ll T$.  

This interpretation of ${d\Gamma(\bm q)/ d^3\bm q}$ can clearly be seen in the heavy quark damping rate, which
is given by the imaginary part of the heavy-quark self-energy from one-gluon loop as in Fig.~\ref{fig:H-L}, that is
the damping rate can be shown to be given by
\be
\gamma^{\rm HQ}=
\int d^3\bm q\,{d\Gamma(\bm q)\over d^3\bm q} \,,
\ee
with the same definition of ${d\Gamma(\bm q)/ d^3\bm q}$ as above. We have
$\rho_L=-2\,{\rm Im}G_R^{00}$ and  
the expression (\ref{scatt}), by cutting
rules, describes the Coulomb scattering between thermal particles and
the heavy quark at rest as shown in Fig.~\ref{fig:H-L}.  The one-gluon
mediation is generally screened by thermal self-energies (blobs in
Fig.~\ref{fig:H-L}) to have IR divergences tamed.  The screening is
provided by contributions of both the LLL quark states and the HTL
gluons,
which will be detailed in the next section.  In the case of
$eB\gg T^2$ the LLL contribution to the screening mass
($m_D^2\sim \alpha_s eB$) will dominate over the gluonic contribution
($\sim \alpha_s T^2$).  As a wrap-up, we emphasize that the time-like
region, 
$\omega^2- \vert \vq\vert^2 > 0$, of the spectral density
$\rho_L(\omega,\bm q)$ does not contribute to the momentum diffusion
coefficients or the scattering rate \eqref{scatt} in the $\omega \to
0$ limit.


\section{Formalism for computation of scattering rates}
\label{sec:computation}

As discussed in the previous section, to compute the heavy quark
diffusion and drag coefficients at LO, we need the
longitudinal spectral density $\rho_L(\omega,\bm q)$ near
$\omega\to 0$, which can be obtained from the retarded gluon correlator
$G_R^{00}(\omega,\bm q)$. To include both the screening effects from and the scatterings with thermal particles, the one-loop gluon self-energy is re-summed into the longitudinal propagator $G_R^{00}(\omega,\bm q)$. Roughly speaking, the real part of this self-energy gives the screening effects, while the imaginary part is responsible for the spectrum of scattering particles. Fermions (i.e. light quarks) and hard thermal
gluons will contribute to the self-energy, and we denote them as
$\Pi^{\mu\nu}_{R, \fermion}$  and $\Pi^{\mu\nu}_{R,\gluon}$,
respectively, throughout this paper.  We first present our computation
for $\Pi^{\mu\nu}_{R,\fermion}$ in the LLL approximation as shown in
Fig.~\ref{fig:polarization}.  We investigate general features of
the gluon self-energy due to the polarization of quarks and antiquarks
in the LLL states in the next subsection~\ref{sec:sf}
and express the resulting spectral density from it in subsection~\ref{sec:G00}.  The thermal
gluon contribution to the self-energy, $\Pi^{\mu\nu}_{R,\gluon}$, will be addressed later in
section~\ref{sec:gluon}.

\subsection{Gluon self-energy from quark loop}
\label{sec:sf}

The two-point functions and the self-energy are diagonal in color
indices, and so we factor them out for notational brevity.  After
taking the color trace, contributions from a particle species in the
representation $R$ to the one-loop self-energy is proportional to the
following pre-factor,
\be
 T_R\equiv {C_R^{\rm LQ}\cdot {\rm Dim}(R)\over {\rm Dim}(G)}\,,
\ee
where ${\rm Dim}(G)=N_c^2-1$ is the dimension of the adjoint
representation and $C_R^{\rm LQ}$ is the light-quark Casimir.  We have
$T_R=1/2$ and $N_c$ for the fundamental and the adjoint
representations, respectively.  These factors take care of the color
representation of light particles inside the loop.

We will utilize the real-time Schwinger-Keldysh formalism in
``ra''-basis.  In this language, we can express the 
retarded gluon self-energy as
\be
 \Pi^{\mu\nu}_{R,\fermion}(Q) = i 4\pi\alpha_s T_R
  \langle J^\mu_r(Q) J^\nu_a(-Q)\rangle \,,
\label{selfe}
\ee
where $J^\mu_{r,a}$ is the current operator (with color indices
amputated as described before) and the subscript ($r$,$a$) refers to
the Keldysh basis.
\begin{figure}
     \begin{center}
              \includegraphics[width=0.8\hsize]{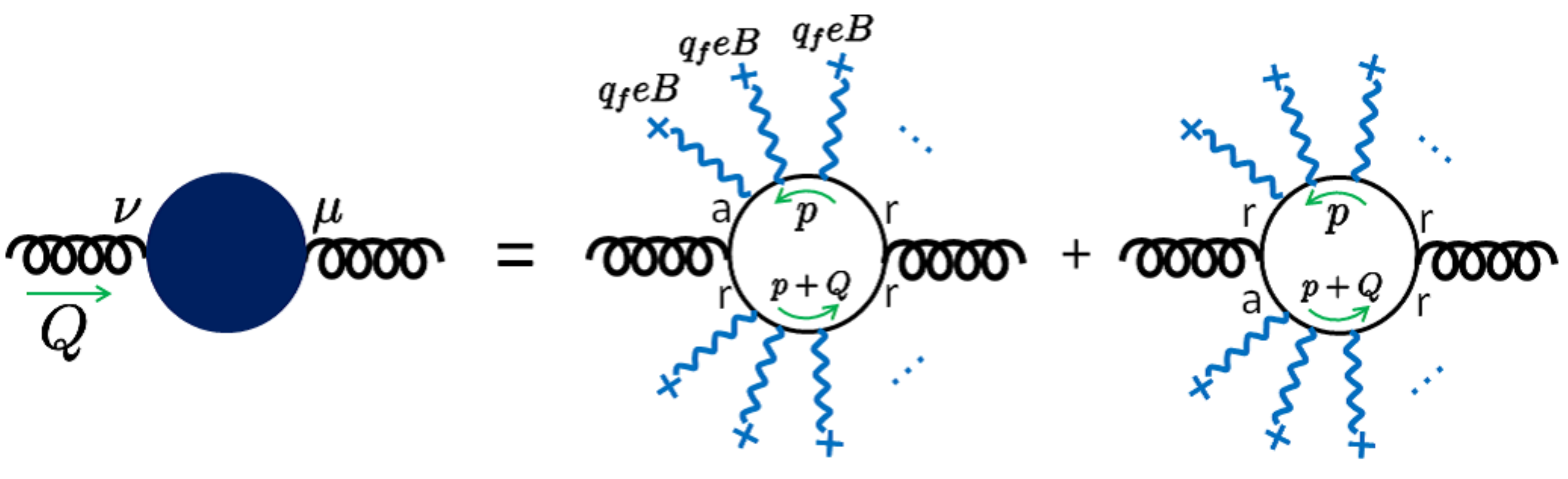}
     \end{center}
\vspace{-0.5cm}
\caption{One-loop gluon self-energy due to the polarization 
of a pair of quark and antiquark LLL states.  Indices ``a'' and ``r''
denote the Schwinger-Keldysh basis.}
\label{fig:polarization}
\end{figure}
Re-summing
insertions of the external magnetic field (see Fig.~\ref{fig:polarization}), 
the real-time quark propagators at finite $T$ in 
the LLL approximation are given by (see, e.g., Ref.~\cite{Fukushima:2011nu} 
for an explicit construction of the quark propagator with B)
\begin{align}
\label{eq:Sra}
 S_{ra}(p) &= i\, e^{-\bm p_\perp^2/|q_f e B|}\,
  {2(\slashed p_\parallel + \mq){\cal P}_-\over p_\parallel^2-\mq^2}
  \bigg|_{p^0\to p^0+i\epsilon}\,,
\\
\label{eq:Sar}
 S_{ar}(p) &= i\, e^{-\bm p_\perp^2/|q_f e B|}\,
  {2(\slashed p_\parallel+\mq){\cal P}_-\over p_\parallel^2-\mq^2}
  \bigg|_{p^0\to p^0-i\epsilon} 
\,,
\\
\label{eq:Srr}
 S_{rr}(p) &= \biggl[{1\over 2}-n_F(p^0)\biggr]\bigl[S_{ra}(p)
  -S_{ar}(p)\bigr] \equiv \biggl[{1\over 2}-n_F(p^0)\biggr]\rho_F(p) \,,
\end{align}
where $q_f$ is the electric charge of quark species $f$ in unit of $e$, and the magnetic field is assumed to be oriented in the z-direction. 
The metric in the longitudinal two-dimensional subspace is defined by $g_\parallel^{\mu\nu} = {\rm diag}\, (1,0,0,-1)$, 
so that $p_\parallel^\mu = g_\parallel^{\mu\nu} p_\nu$ and 
$\slashed p_\parallel=p^0\gamma^0-p^z\gamma^z$.  
The spin-projection operators are defined by
${\cal P}_\pm \equiv (1\pm i {\rm sgn}(q_f B) \gamma^x\gamma^y)/2$, which project quark fields onto the (1+1)-dimensional spinors.
Note that these operators depend on the quark charge $q_f$ since the spin magnetic moment depends on the quark charge. 
Nevertheless, our final result for $\kappa$ will be independent of the sign of $q_f B$
because of the charge-conjugation invariance.
The bare quark spectral density, from Eqs.~(\ref{eq:Sra}) and (\ref{eq:Sar}), is given by 
\be
 \rho_F(p) = e^{-\bm p_\perp^2/|q_feB|}( \slashed p_\parallel +\mq)
  \,{\cal P}_-\, {2\pi\over p^0}\Bigl[
  \delta\Bigl(p^0-\varepsilon_{p_z}\Bigr)
 +\delta\Bigl(p^0+\varepsilon_{p_z}\Bigr) \Bigr] 
\, ,
\ee
where the dispersion relation for the LLL states, $\varepsilon_{p_z} = \sqrt{p^2_z+\mq^2}$, is purely two dimensional and
is independent of $q_f$. 


Figure~\ref{fig:polarization} shows two diagrams contributing to Eq.~\eqref{selfe} in the real-time ra-basis,
which yields
\be
 \langle J^\mu_r(Q)J^\nu_a(-Q)\rangle = -\int {d^4 p\over (2\pi)^4}
 \,{\rm tr}\Bigl[ \gamma^\nu S_{ar}(p)\gamma^\mu S_{rr}(p+Q) +\gamma^\nu S_{rr}(p)\gamma^\mu S_{ra}(p+Q)
  \Bigr] \,.
\label{loop}
\ee
The first thing to notice is that the
integration with respect to $\bm p_\perp$ is trivially factorized as
\be
\label{transv}
 4\int{d^2\bm p_\perp\over (2\pi)^2}\;
  e^{-\bm p_\perp^2/ |q_f eB|} \, e^{-(\bm p_\perp+\bm q_\perp)^2/ |q_f e B|}
 = { |q_f e B|\over 2\pi}\; e^{-\bm q_\perp^2/ (2 |q_f eB|)} 
\,.
\ee
This is naturally expected since the transverse dynamics decouples
from the longitudinal dynamics of the LLL states; the operator
${\cal P}_-$ projects the (3+1)-dimensional (Dirac) fermions onto
(1+1)-dimensional ones. The factor of $|q_f eB|/(2\pi)$ can be understood as
the transverse density of states for the LLL states.
Moreover, from the fact that the transverse $\gamma$-matrices (namely,
$\gamma^x$ and $\gamma^y$ collectively denoted by $\gamma^\perp$)
satisfy
\be
 {\cal P}_\pm \gamma^\perp = \gamma^\perp {\cal P}_\mp \,,
\ee
we see that the transverse components of the self-energy are zero,
i.e.\ $\Pi_R^{\perp \mu}=\Pi_R^{\mu\perp}=0$.  This is physically
clear from the absence of transverse currents with the LLL states.
The remaining longitudinal part of the self-energy after performing
the integral over $(p^0,p_z)$ in (\ref{loop}) should be equivalent to the
corresponding one for the (1+1)-dimensional Dirac fermion at finite temperature.

We thus have, after summing over quark flavors $f$,
\begin{equation}
\label{eq:dim_reduction}
\Pi^{\mu\nu}_{R,\LLL}(Q)=  \pi s(\vq_{\perp})\,\Pi^{\mu\nu}_{R, {\rm 2D}}(\o, q_z)\, ,
\qquad
s(\vq_{\perp})\equiv 4\alpha_{s} T_{R} 
\sum_f \(\frac{|q_f eB|}{2\pi}\) e^{-\frac{\vq^{2}_{\perp} } {2 |q_f eB|}}
 \, ,
\end{equation}
where $\Pi^{\mu\nu}_{R,{\rm 2D}}(\o, q_{z})$ is 
the retarded self-energy tensor in $1+1$~dimensions 
which is dimensionless and is independent of $q_f B$. 
The gauge-invariance completely fixes its form
as $\Pi^{\mu\nu}_{R,\,{\rm 2D}}(\omega, q_{z}) \propto
\bigl( q_\parallel^2 g_\parallel^{\mu\nu}-q_\parallel^\mu q_\parallel^\nu \bigr)$, so that
one can write $\Pi^{\mu\nu}_{R,\LLL}$ as
 \begin{equation}
 \label{Pi_bar_def}
 \Pi^{\mu\nu}_{R, \LLL}(\o, \vq) \equiv
 \bar{\Pi}_{\LLL}(\o, \vq)\bigl( q_\parallel^2 g_\parallel^{\mu\nu}-q_\parallel^\mu q_\parallel^\nu \bigr) \,,
 \end{equation}
which defines $\bar{\Pi}_{\LLL}(\o,\vq)$.  Note that this is the unique gauge-invariant tensor
 structure in (1+1) dimensions, independent of whether (1+1)-dimensional Lorentz symmetry is broken at finite temperature.

From Eq.~\eqref{eq:dim_reduction} we see that the self-energy from the
LLL states is of the order of $ \alpha_s eB$.  On the other hand, the
contributions from hard thermal gluons to the self-energy, for example
to the screening mass $m_D^2$, is of order $\alpha_s T^2$, which is
sub-dominant compared to the LLL contribution by $eB\gg T^2$.  We
can therefore neglect thermal gluon contributions to the self-energy
up to this order.  This defines our LO computation in the
regime of our interests, $\alpha_s eB \ll T^2\ll eB$.  In section~\ref{sec:massless} we
will find, however, that the longitudinal momentum diffusion
coefficient $\kappa_\parallel$ vanishes in this LO approximation in
the massless limit (see the
next section for more details), which necessitates including the
imaginary part of the thermal gluon contribution to the self-energy to find a non-zero contribution to $\kappa_\parallel$.
Then, in this way, we can get a first leading non-zero value of
$\kappa_\parallel$ in the $\mq=0$ limit, which is suppressed by a
power of $T^2/(eB)$ as compared to $\kappa_\perp$ as elucidated in
section~\ref{sec:gluon}.  Thus, our definition of ``LO'' for $\kappa_\parallel$ in the
$\mq=0$ case should be understood in this sense.  We emphasize that
our computation for $\kappa_\parallel$ in the massless limit at this
modified LO is also systematic and well-defined in accord with
$\alpha_s eB\ll T^2\ll eB$, as will be explained in section~\ref{sec:gluon}.

\subsection{Scattering rate from the spectral density}
\label{sec:G00}

Based on the above discussion, let us temporarily neglect thermal gluon
contributions to the self-energy for the moment, and the tensor structure of the gluon self-energy from the LLL states is then
given in Eq.~\eqref{Pi_bar_def}.  We shall first elaborate the tensor
structure of the resulting gluon two-point correlation function with the above self-energy. 

In general the expression for the $A^0$ gluon propagator,
$G_R^{00}$, depends on the gauge choice.  Nevertheless, the final
expression for the Coulomb scattering rate of a static heavy quark is physical and is
gauge invariant.  A simple way to see the gauge invariance is to note
that the gauge transformation of $A^0$ is of a form,
$A^0\to A^0+\omega\alpha$, and since $\langle A^0\rangle=0$, the
correction to $G_R^{00}$ is of order $\omega^2$, which vanishes in Eq.(\ref{scatt}) for the scattering rate in
$\omega\to 0$.  We will demonstrate the gauge invariance with two choices of gauge fixing;  the
covariant and the Coulomb gauges.  Inserting 
the self-energy~\eqref{Pi_bar_def}, we can write down the general form
of the gluon retarded propagator in the covariant gauge as~\cite{Hattori:2012je}
\be
\label{cov}
 G^{00}_R(Q) = 
 \frac{\omega^2}{Q_\epsilon^4}
 \bigg ( \frac{\vq^2_{\perp}}{q^2_{\parallel}} - \xi\bigg)
  + \frac{q^2_{z}}{q^2_{\parallel}}\, \biggl[
  \frac{1}{Q^2-q_\parallel^2\bar{\Pi}_{LLL}(\omega+i\epsilon,\bm q)} \biggr] \, ,
\ee
where $Q_\epsilon^2\equiv(\omega+i\epsilon)^2-\bm q^2$ and $\xi$ is a
gauge parameter.  Note that we do not insert $i\epsilon$ in $Q^2$ and $q_\parallel^2$ that appear in the above, especially in the second part, since they come from the tensor structure in Eq.~\eqref{Pi_bar_def} and an $i\epsilon$ in
$\bar{\Pi}_{LLL}$ is sufficient to keep the imaginary part correctly.
On the other hand, in the Coulomb gauge we have
\be
 G^{00}_R(Q) = {1\over \bm q^2}\,{1\over Q^2-q_\parallel^2
  \bar{\Pi}_{LLL}(\omega+i\epsilon,\bm q)}
  \Biggl[ Q^2+{(Q^2-q_\parallel^2) \omega^2
  \bar{\Pi}_{LLL}(\omega+i\epsilon,\bm q)\over
  \bm q^2}\Biggr] 
- \xi\frac{\omega^2}{|\boldsymbol{q}|^4}  
  \,.
\label{coul}
\ee
We can readily confirm that the above two expressions give
an identical form for the scattering rate in the $\omega\to0$ limit, i.e.,
\begin{equation}
\label{GR_LLL}
\frac{d\Gamma(\vq)}{d^3\vq}={4\pi\alpha_s \over (2\pi)^3}C_R^{\rm HQ} \lim_{\omega\to0}{T\over\omega}(-2){\rm Im}[G^{00}_R(\omega,\bm q)]
=\frac{\alpha_s T}{\pi^2}C^{HQ}_{R}\, \frac{f_{\LLL}(\vq)}{\[\vq^2+\Re\, \Pi^{00}_{R,LLL}(\o=0,\vq)\]^2}\, ,
\end{equation}
where we introduced $\Im\, \Pi^{00}_{R, \LLL}(\o, \vq)=\o f_{\LLL}(\vq)$
for small $\o$ as it is an odd function of $\omega$ in general, and
we have used Eq.~\eqref{Pi_bar_def} to find 
$\Pi^{00}_{R, \LLL}(0, \vq)=-q_z^2\bar{\Pi}_{\LLL}(0, \vq)$.

Before moving to the computation of $\Pi^{00}_{R, \LLL}(Q)$ which will
be addressed in the next subsection, it would be instructive to recall
the well-known picture of heavy quark scatterings (without a strong
magnetic field), and compare it with our case of the LLL states.
Without magnetic field, the imaginary part of $G^{00}_{R,B=0}$ comes
from the imaginary part of $\Pi^{00}_{R,B=0}$ in the covariant gauge
(which is an odd function of $\omega$) or more precisely expressed
as $F(Q)\equiv (-Q^2/\bm q^2)\Pi_{R,B=0}^{00}$ in the common
convention of Ref.~\cite{LeBellac}.  As discussed shortly, from the
HTL contribution to
$\Pi^{00}_{R,B=0}$ for soft $Q$ (i.e.\ from hard thermal gluons and quarks), the resulting
spectral density $\rho_{L,B=0}(Q)=-2{\rm Im}[G^{00}_{R,B=0}(Q)]$
receives the following two contributions:  (i) The plasmon pole located in
a time-like region ($|\omega|>|\bm q|$).  (ii) The continuous part
along the space-like interval $(|\omega|<|\bm q|)$ originating from
the Landau damping.

The plasmon pole remains gapped even in the $|\bm q|\to 0$ limit by
the plasma frequency, $\omega^2\to m_{\rm pl}^2=m_{D,B=0}^2/3$, with the
Debye mass
$m_{D,B=0}^2\equiv \Re\Pi_{R,B=0}^{00}(\omega=0)=(4\pi\alpha_s/3)(N_c+T_{R}N_f)T^2$.
Thus, in our limit of $\omega\to 0$, the plasmon pole (i) decouples
and only the 
continuous Landau damping part (ii) is relevant.  In this
case, we can smoothly take the $\omega\to 0$ limit for
$\rho_L(Q)/\omega$, that is,
\be
\label{eq:ReIm}
 \lim_{\omega\to 0}(-2) {\Im [{G_{R,B=0}^{00}(Q)}]\over\omega}
  = \lim_{\omega\to 0}
  \frac{2}{\omega}\,\frac{\Im\Pi_{R,B=0}^{00}(Q)}
  {[Q^2-\Re\Pi_R^{00}(Q)]^2+[\Im\Pi_{R,B=0}^{00}(Q)]^2}
 = {2 f(\bm q)\over (\bm q^2+m_{D,B=0}^2)^2} \,,
\ee
where we can write
$\Im\Pi_{R,B=0}^{00}(\omega\sim 0)=\omega f(\bm q)$ because $\Im\Pi_{R,B=0}^{00}(\omega)$ is an odd function of
$\omega$\footnote{More explicitly, we have $f(\bm q)=\pi m_D^2/2|\bm q|$ for
  soft $\bm q^2\sim \alpha_sT^2$, while for ultra hard $|\bm q|\gg T$,
  it becomes the Boltzmann factor suppressed as $f(\bm q)\sim e^{-|\bm q|/T}$.  Also, it
  can be shown that $\Re[\Pi^{00}_{R,B=0}(\omega=0,\bm q)]$ behaves as
  $\sim \alpha_s T^4/\bm q^2$ for ultra hard $|\bm q|$.},
and thus we find a finite scattering rate. 
From the cutting rules, the physics
interpretation is clear;  $f(\bm q)$ is an integrated spectrum of
scattering particles of momentum transfer $\bm q$ weighted by thermal
distributions, while $1/(\bm q^2+m_{D,B=0}^2)^2$ is the square of the
screened Coulomb amplitude.

We find that a similar physics interpretation is also possible
for the spectral density from the LLL contributions in (\ref{GR_LLL}).  There exists a
time-like plasmon pole determined by 
\begin{equation}
\label{plasmon}
 Q^2=q^2_{\parallel}\, \Re\,\bar{\Pi}_{LLL}(\o, \vq)
\end{equation}
with a mass of order $m_{\rm pl}^2\sim \alpha_s eB$, and the spectral
density from this pole is irrelevant in the $\omega=0$ limit.  In the
space-like region (and on the light-cone in the massless limit),
especially near $\omega=0$, we will explicitly show that there exists
a finite spectral density coming from the Landau damping with the LLL
states.  In fact, Eq.~\eqref{GR_LLL} has precisely the same structure as in
 Eq.~\eqref{eq:ReIm}.
As mentioned below Eq.~\eqref{eq:ReIm}, the factor $f_{\LLL}(\bm q)$ from the
imaginary part again represents the integrated spectrum of the
scatterers, namely the quarks thermally populated in the LLL states.
On the other hand, the real part in the denominator of
Eq.~\eqref{GR_LLL} provides a screening for the Coulomb scattering
with the screening mass, $m_{D,B}^2\sim \alpha_s eB$. We will find
that the LO contributions to $\kappa_{\perp}$ come from the momentum
transfer region $\sqrt{\alpha_s eB}\lesssim |\bm q_\perp|\lesssim \sqrt{eB}$
as we mentioned below Eq.~\eqref{eq:fdtheorem}.  The upper
cutoff, $\sqrt{eB}$, arises from
the Gaussian factor of the quark propagator,  and it reflects the fact that the LLL states carry intrinsic transverse
momentum of order $\sim \sqrt{eB}$ even at $T=0$ due to their transverse size $l\sim 1/\sqrt{eB}$, and the transverse
momentum transfer is bounded not by thermal distribution of the
scattering LLL particles (for which the energy levels are independent
of $q_\perp$) but by $\sqrt{eB}$.  This explains why the upper cutoff
is not given by $T$ from the Boltzmann factor $e^{-|\bm q|/T}$, which would normally be the case in other finite-$T$
calculations.

\section{Massless limit}
\label{sec:massless}

In this section we consider the case where the light quarks (of
representation $R$) in the LLL states are massless, i.e.\ $\mq=0$.
We will find some special features originating from the nature of
chiral fermions.  To evaluate the spectral density in
Eq.~\eqref{GR_LLL}, we need to determine $\Pi^{00}_{R,\LLL}$ which,
according to Eq.~\eqref{eq:dim_reduction}, is cast into the problem of
computing $\Pi^{\mu\nu}_{R,\rm 2D}(\o, q_z)$ from the massless fermion one-loop in tow dimensions (2D).  
In this case we can use a powerful technique of bosonization
that maps (1+1)-dimensional massless 
fermions into bosons~\cite{Mandelstam:1975hb}.  The mapping rule is
well established as (see, for example, Ref.~\cite{Smilga:1992hx} and
also Ref.~\cite{Fukushima:2011jc} for the application to QCD in strong
magnetic field)
\be
J^\mu_{\rm 2D}=\sqrt{1\over\pi}\,\epsilon^{\mu\nu}\partial_\nu\phi\,,\qquad
J^{A,\mu}_{\rm 2D}=\sqrt{1\over\pi}\,\partial_\mu\phi\,,
\ee
between the vector (axial) current $J^\mu_{\rm 2D}$ ($J^{A,\mu}_{\rm 2D}$) and a real scalar field $\phi$.  We note that
\be
 \langle \phi_r(q_\parallel)\phi_a(-q_\parallel)\rangle
   = {i\over (\omega+i\epsilon)^2-q_z^2}
\ee
for any temperature $T$ and chemical potential $\mu$, since the retarded two point function of a free theory is independent of $(T,\mu)$.  
Using this correspondence, we can easily get the retarded
current-current correlator in (1+1) dimensions as
\begin{align}
\Pi^{\mu\nu}_{R,{\rm 2D}}(\omega,q_z) &\equiv
 i\langle J^\mu_r(q_\parallel) J^\nu_a(-q_\parallel)\rangle_{\rm 2D}
=
\frac{i\epsilon^{\mu\alpha}\epsilon^{\nu\beta}q_{\parallel\alpha}q_{\parallel \beta} }{\pi} \,
 \langle \phi_r(q_\parallel)\phi_a(-q_\parallel)\rangle
\notag\\
  &= \frac{1}{\pi \[(\omega+i\epsilon)^2-q_z^2\]} 
   \Bigl(q^2_{\parallel}g^{\mu\nu}_\parallel
  -q^\mu_\parallel q^\nu_\parallel\Bigr)
\,.
\label{Pi_2D}
\end{align}
We have explicitly checked the cancellation of all
$T$ dependent terms in a direct computation of the fermion loop in
Eq.~\eqref{loop} at finite $T$.  Consequently, comparing
Eqs.~\eqref{eq:dim_reduction} and \eqref{Pi_bar_def} with
Eq.~\eqref{Pi_2D}, we find
\begin{subequations}
\label{eq:PiLLL}
\begin{align}
\Re\, \Pi^{00}_{R,\LLL}(\o, \vq) &=
 - s(\vq_{\perp})\frac{q^2_{z}}{q^2_{\parallel}}\, , 
    \\
\Im\, \Pi^{00}_{R,\LLL}(\o, \vq) &=
{ \pi\over 2} s(\vq_{\perp}) \, \o\,  \bigl[ \delta(\o-q_{z})+\delta(\o+q_{z})\bigr]\, . 
\end{align}
\end{subequations}
 
As we discussed previously, the spectral density inferred from the
imaginary part of $G^{00}_R(Q)$ has two pieces:
a plasmon pole and a continuous part from the Landau
damping\footnote{There is a $Q^2=0$ pole in the covariant gauge that
does not exist in the Coulomb gauge, and one may wonder how they can
be consistent.  This difference is well-known.  As explained in
Ref.~\cite{Kapusta:2006pm}, the Coulomb gauge has two polarizations
corresponding to the physical modes, while the covariant gauge has
extra two unphysical modes.  Those additional degrees of freedom give
rise to $Q^2=0$ pole in such a way not to affect physical observables,
and so we can discard this pole safely.}.
It is easy to find that Eq.~\eqref{plasmon} gives rise to a time-like
plasmon with the following dispersion relation,
\be
\label{plasmon_zero}
 \omega^2 
 = \bm q^2 +s(\vq_{\perp})
 = \bm q^2
+ 4\alpha_{s} T_{R} \sum_f \(\frac{|q_f e B|}{2\pi}\) e^{-\frac{\vq^{2}_{\perp} } {2 |q_f e B|}}
\, ,
\ee
where we have used the real part shown in Eq.~\eqref{eq:PiLLL}.  This
is nothing but a plasmon carrying a mass of order of
$m_{\rm pl}^2\sim \alpha_s eB$ for $\bm q_\perp^2\lesssim eB$.  This
plasmon mass, that exists even in the $\mq=0$ limit, can be
interpreted as a result of Schwinger's anomalous mass generation in
(1+1)-dimensional massless gauge theory~\cite{Schwinger:1962tn}, which
is in the present case extended to a theory in (3+1) dimensions with
the overall transverse Landau degeneracy factor.  It is clear that
this time-like plasmon is gapped and it produces no contribution to
the static spectral density.

Now, a finite contribution to the Coulomb scattering is obtained by
inserting Eq.~\eqref{eq:PiLLL} into Eq.~\eqref{GR_LLL} as
\begin{equation}
\label{eq:spect_massless}
\frac{d\Gamma(\vq)}{d^3\vq}
=\frac{\alpha_s T}{\pi}C^{HQ}_{R}\, \frac{s(\bm q_\perp)}{\[\vq^2+s(\bm q_\perp)\]^2}\delta(q_z)\, ,\end{equation}
where the static limit $\omega\to0$ in Eq.~\eqref{eq:PiLLL} results in
the delta function of $q_z\to0^\pm$, and we repeat the definition of $s(\bm q_\perp)$ 
\be
\quad
s(\bm q_\perp)=4\alpha_{s} T_{R} \sum_f \(\frac{|q_f e B|}{2\pi}\) e^{-\frac{\vq^{2}_{\perp} } {2 |q_f e B|}}\,.
\ee
 This is a central result in this
section. One can easily compute the leading order heavy quark damping rate from this.

This delta function $\delta(q_z)$ associated with the LLL states can
be understood from a simple kinematic constraint with
(1+1)-dimensional massless fermions.  These fermions have dispersion
relations, $E(p)=\pm p_z$, where the sign refers to the chirality.
Since perturbative QCD interactions do not flip the chirality, the
energy-momentum transfer from these massless fermions,
$(\Delta E,\Delta p_z)=(\omega,q_z)$, must satisfy $\omega=\pm q_z$.
Then, the static limit $\omega\to 0$ imposes the vanishing
longitudinal momentum transfer, which is represented by the delta
function $\delta(q_z)$.

First, we immediately conclude from Eqs.~\eqref{kappa}
and \eqref{eq:spect_massless} that the longitudinal momentum diffusion
coefficient
from the LLL states is strictly zero because of the vanishing
longitudinal momentum transfer constrained by the kinematics in the
massless case at LO.  In contrast to this, as we
will see in section~\ref{sec:massive}, a finite light quark mass
introduces $\kappa_\parallel\propto \mq^2\neq 0$ at LO. We will
also see in section~\ref{sec:gluon} that $\kappa_\parallel$ 
acquires a non-zero contribution from the scatterings with the hard thermal gluons (but the exchanged gluon is still
screened by the LLL states at LO). The resulting $\kappa_\parallel$ is smaller than $\kappa_\perp$ by a factor of $T^2/eB$.

\begin{figure}
     \begin{center}
              \includegraphics[width=0.6\hsize]{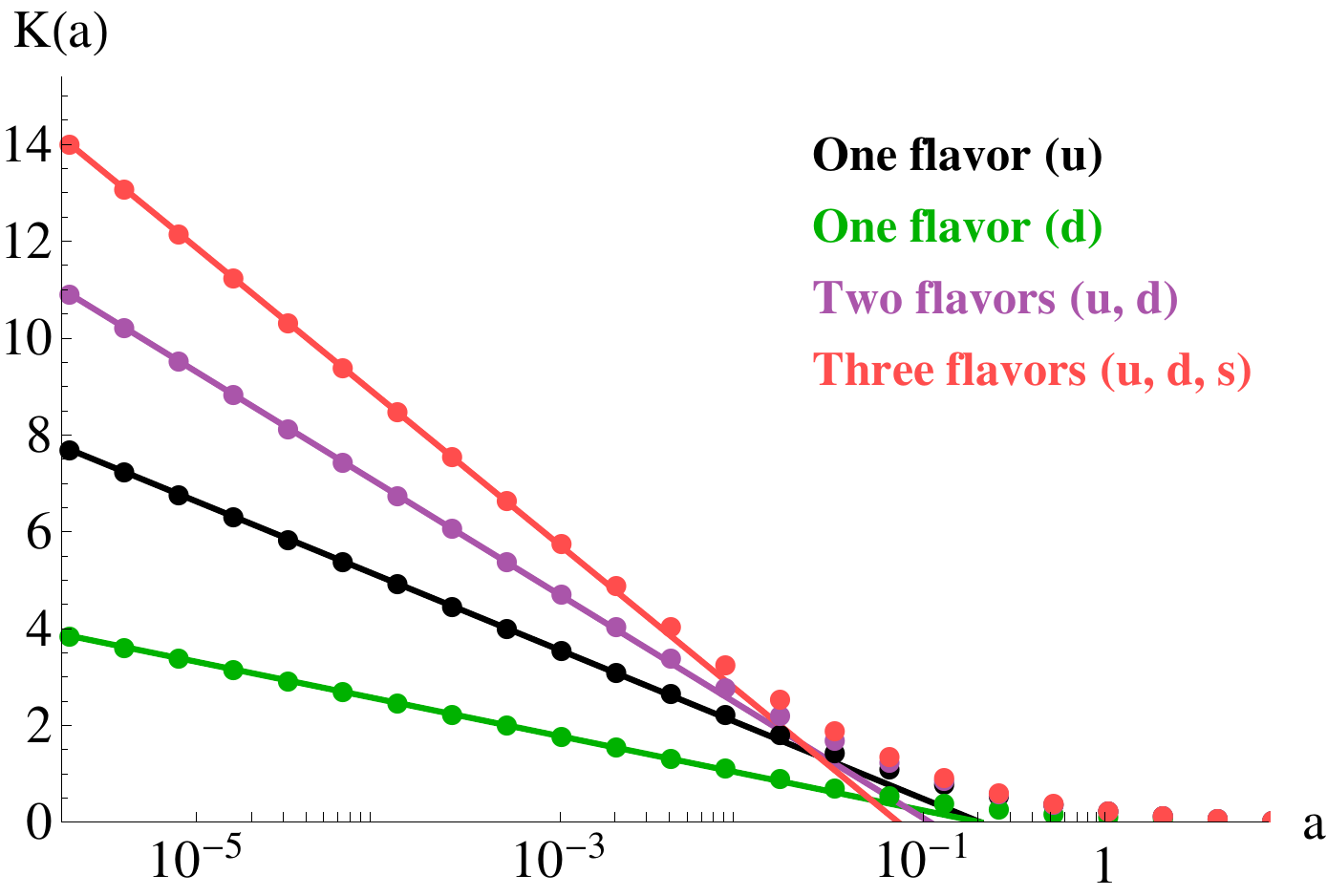}
     \end{center}
\vspace{-0.5cm}
\caption{
Integral (\ref{eq:integ}) for the transverse momentum diffusion coefficient. 
Lines show the analytic expressions given in between the brackets in Eqs.~(\ref{LO}), 
which are confirmed by the numerical integration shown by filled circles. }
\label{fig:kappa_perp}
\end{figure}

From the above leading order scattering rate, we can finally obtain the
transverse momentum diffusion coefficient as
\be
\kappa_\perp = \frac{\alpha_s T}{2\pi}  C_R^{\rm HQ} \int \!\! d^2 q_\perp 
\vert \vq_\perp \vert^2 \frac{ s(\bm q_\perp) }
{ \left[ \, \vert \bm q_\perp \vert^2+s(\bm q_\perp) \, \right]^2}
 = 2\alpha_s^2 T C_R^{\rm HQ} T_R  \biggl({eB\over 2\pi}\biggr) K(a)
\, ,
\label{LOinteg}
\ee
where we defined an integral as 
\begin{eqnarray}
K(a) = \int_0^\infty dx\,
  {x\, \sum_f  |q_f| e^{-x/{ |q_f|}  } \over [x+ a \, \sum_f |q_f| e^{-x/{|q_f|}} ]^2 }
\label{eq:integ} 
\, ,
\end{eqnarray}
with dimensionless variables $ x\equiv |\bm q_\perp|^2 / 2eB$ and 
$a \equiv \alpha_s T_R / \pi$. 
It is easy to see that the leading log contribution in $\alpha_s$ comes from the range
$\alpha_s\lesssim x\lesssim 1$ or equivalently
$\sqrt{\alpha_s\,eB}\lesssim |\bm q_\perp|\lesssim \sqrt{eB}$ as
pointed out before.  In this range, the integrand is approximately
$1/x$, and the integration produces the leading log behavior as $\sim \log(1/\alpha_s)$.
A more careful evaluation gives the full LO result including the constant under the logarithm as 
\begin{eqnarray}
 \kappa_\perp^{\rm LO} = 2\alpha_s^2 T C_R^{\rm HQ} T_R 
  \biggl({eB\over 2\pi}\biggr) \cdot Q_{\rm em}  \biggl[
\log\biggl({1\over \alpha_s}\biggr) 
  - \log\biggl({T_R \over\pi}\biggr) - \gamma_E - 1 
+ \sum_f \frac{|q_f|}{Q_{\rm em} } \log(\frac{|q_f|}{Q_{\rm em}}) \biggr]
\, ,
\nonumber
\\
\label{LO}
\end{eqnarray}
where $\gamma_E\approx 0.577$ is the Euler-Mascheroni constant, 
and the sum of electric charges is defined by $Q_{\rm em} = \sum_f |q_f|$. 
As shown in Fig.~\ref{fig:kappa_perp}, 
we have numerically checked that this final form is a good
approximation to Eq.~(\ref{LOinteg}) as long as
$a \sim \alpha_s \ll 1$.

Let us briefly discuss contributions of the higher Landau levels
(hLLs) which are suppressed in both the vacuum and the thermal parts
of the gluon self-energy in the strong $B$ limit.  First, we should
note that the momentum transfer $q$ corresponds to the external
momentum of the gluon self-energy and that only the space-like
momentum transfer $(q_\parallel^2 - q_z^2 < 0)$ contributes to the
heavy quark momentum diffusion.  In the vacuum part, contributions
from the hLLs are suppressed because quarks and antiquarks have the
dispersion relation, $p_\parallel^2 = \mq^2 + 2neB $ $(n\geq 1)$, and
the off-shellness is of order of $eB$, which is because the momentum
$q_\parallel^2$ is located away from the pair creation threshold
(i.e.\ on-shell point) by $eB$~\cite{Hattori:2012je}.  Therefore, one
can conclude that contributions from the hLLs are suppressed at least
by ${\mathcal O(1/eB)}$ when either a quark or an antiquark is excited
to a hLL, and by ${\mathcal O(1/eB^2)}$ when both of them belong to
hLLs.  As for the thermal part, contributions from the hLL are
exponentially suppressed by the Boltzmann factor $\sim
e^{- \sqrt{2neB}/T}$.

\section{Finite contributions to the longitudinal momentum diffusion coefficient}
\label{sec:subleading}

As we have seen in the previous section, the longitudinal momentum
diffusion coefficient vanishes when we consider only the massless
quarks in the LLL states.  This is a consequence of
the massless (1+1)-dimensional dispersion relation of the LLL states,
which does not allow for any longitudinal momentum transfer at
$\omega=0$.  In this section, we examine light-quark mass corrections
and thermal gluon contributions.

\subsection{Light-quark mass effects}
\label{sec:massive}

In this subsection we consider finite mass corrections to
$\kappa_\parallel$, which can relax the constraint of the
longitudinal momentum transfer that is strictly prohibited in the
massless case.  First we emphasize that the basic
structure of the gluon self-energy shown in
Eqs.~\eqref{eq:dim_reduction} and \eqref{Pi_bar_def} are still valid
regardless of the quark mass.  Then, one will immediately find that
the expression of the spectral density~\eqref{GR_LLL} is also
intact and that the problem reduces to computation of 
$\Pi^{\mu\nu}_{R,{\rm 2D}}(\omega, q_{z})$ that should replace the
massless~\eqref{Pi_2D} with the massive one.

We should notice that the transverse dynamics of LLL quarks are
not directly affected by mass corrections as clearly seen in the
propagators~\eqref{eq:Sra}--\eqref{eq:Srr}.  Therefore, the gluon
self-energy can be written in the same form as
Eq.~\eqref{eq:dim_reduction}:
\begin{align}
\Pi^{\mu\nu}_R (\o, q_{z})
&= 
\pi \, s(\vq_\perp)
\Pi_{R,{\rm 2D}}(\o, q_{z})
\bigl( q_\parallel^2 g^{\mu\nu}_\parallel - q_\parallel^\mu q_\parallel^\nu\bigr)
\, ,
\label{eq:mass}
\\
\Pi_{R,{\rm 2D}}(\o, q_{z}) &=
\Pi^{\rm \vac}_{R,{\rm 2D}}(q_\parallel^2) + \Pi^{\rm \th}_{R,{\rm 2D}}(\omega,q_z)
\,.
\end{align}
This is due to the fact that there is only one gauge-invariant
tensor structure in (1+1) dimensions, so that the tensor structure in
Eq.~\eqref{Pi_bar_def} also persists without modification at finite
$T$.  
Since we have an overall factor of $eB$, the
coefficient functions, $\Pi^{\rm \vac}_{R,{\rm 2D}}(q_\parallel^2)$
and $\Pi^{\rm \th}_{R,{\rm 2D}}(\omega,q_z)$, are dimensionless.  We
note that, while $\Pi^{\rm \vac}_{R,{\rm 2D}}(q_\parallel^2)$ depends
on $q_\parallel^2$ in a boost invariant manner,
$\Pi^{\rm \th}_{R,{\rm 2D}}(\omega,q_z)$ depends on $\omega$ and $q_z$
separately due to finite temperature effects.

First, let us consider the vacuum part.  The vacuum part has been
explicitly computed previously in
Refs.~\cite{Hattori:2012je, Fukushima:2011nu,  Hattori:2012ny} and takes 
the following form
\be
\Pi^{\rm vac}_{R,{\rm 2D}}(q_\parallel^2) = 
\frac{1}{\pi \bigl[(\omega+i\epsilon)^2-q_z^2\bigr]}  \Biggl[
 1 - \frac{(2\mq)^2}{\sqrt{q_\parallel^2 [ (2\mq)^2-q_\parallel^2 ] }}\,
 \arctan \Biggl( \frac{q_\parallel^2}{\sqrt{q_\parallel^2 [ (2\mq)^2-q_\parallel^2]}}
  \Biggr) \Biggr] \,,
\label{eq:vac}
\ee
for $q_\parallel^2 \leq 0$.  We can deduce an expression for
$q_\parallel^2>0$ using the analytic continuation.  The first term
corresponds to the massless Schwinger model as discussed in
section~\ref{sec:massless}, and the mass correction has an overall factor
of $\mq^2$.
Therefore, as expected from the dimensional argument, the mass
correction comes with a function of the dimensionless ratio
$\mq^2/q_\parallel^2$.  We will shortly see that the leading order result comes from the momentum transfer range $|\bm q|^2\sim m_{D,B}^2 \sim \alpha_s eB$, so that $\mq^2/q_\parallel^2\sim \mq^2/(\alpha_s eB)$.
In realistic
situations, the light quark mass, $\mq\sim 5\,\text{MeV}$, is much
smaller than other scales, so that we will explore a specific
regime, $\mq^2\ll \alpha_s eB$, and compute the longitudinal momentum
diffusion coefficient $\kappa_\parallel^{\rm LO,\,massive}$ to the
first non-vanishing order in terms of $\mq^2/(\alpha_s eB)$.  Within
this hierarchy, we can safely neglect the $\mq\neq0$ correction to the
real part of the self-energy, that is $m_{D,B}^2$, which is of order $\mq^2/(\alpha_s eB)$ smaller compared to the massless case.

In contrast to the massless case,
$\Pi^{\rm \vac}_{R,{\rm 2D}}(q_\parallel^2) $ acquires an imaginary
part above the threshold of the pair creation at
$q_\parallel^2 = (2\mq)^2 > 0$.  However, this imaginary part in the
time-like region does not contribute to the heavy quark momentum
diffusion in the static limit.

We next compute the thermal part starting with Eq.~\eqref{loop}, 
which can be finally cast into the following form [introducing a compact
notation $Q=(\omega,q_z)$]:
\be
 \Pi^{\rm th}_{R,{\rm 2D}}(Q) =  \frac{\mq^2}{q_\parallel^2} \biggl[ J_0(Q)
  +2{q_z\over q_\parallel^2} J_1(Q)\biggr]
\label{eq:mass_th}
\,,
\ee
with the definition 
\begin{align}
 J_\beta (Q) &\equiv \int_{-\infty}^{\infty}{d p_z\over 2\pi}
  \biggl[{n_+(\varepsilon_{p_z})+n_-(\varepsilon_{p_z})\over \varepsilon_{p_z}}
  \biggr] {p_z^\beta \over (p_z-{1\over 2}q_z)^2-{\omega^2\over 4q_\parallel^2}
  [q_\parallel^2-(2\mq)^2]-i\omega\epsilon}
  \label{eq:integ_th}
  \, ,
\end{align}
where $\beta=0, \, 1$ and we introduce
$\varepsilon_{p_z}\equiv \sqrt{p_z^2+m_q^2}$ and 
$n_\pm(\varepsilon_p)\equiv [e^{(\varepsilon_p\mp\mu)/T}+1]^{-1}$.  
We simplified the retarded $i\epsilon$-prescription (i.e., $\omega \to \omega + i \epsilon$) in Eq.~\eqref{eq:integ_th} for small
$\o$.  We note that $\Pi_{R,{\rm 2D}}^{\rm th}(Q)$ again has an
overall factor of $\mq^2$ as in the vacuum part.  Therefore, the mass
correction goes like $\mq^2/q_z^2$ and $\mq^2/T^2$ and they are
negligible for the real part of the self-energy or the screening mass.

The only important effect for us is the mass corrections to the
imaginary part of $\Pi^{\rm th}_{R,{\rm 2D}}(Q)$, which appear from the
singularities in the integral at $\omega^2=0$ and $\omega = (2\mq)^2$. 
The imaginary part appearing from the factor of $ 1/q_\parallel^2$ is again 
the contribution of the forward scattering as in the massless case, 
so that it does not contribute to the longitudinal momentum diffusion. 
It is instructive to see another expression which can be obtained 
by taking the static limit $\omega\to0$ 
in Eq.~(17) of Ref.~\cite{Baier:1991gg} (in the absence of charge chemical potential) as 
\be
\label{Pi_2D_m}
 \lim_{\omega\to 0}\biggl[\frac{\Im\, \Pi^{00}_R(Q)}{\o}\biggr]= 
 \frac{1}{2T} \int d k_{z} \biggl(1+\frac{k_zk'_z+\mq^2}{\varepsilon_{k_z}\varepsilon_{k'_z}}\biggr)  n_F(\varepsilon_{k_z})
 \bigl[  1-n_F(\varepsilon_{k_z})\bigr]\delta (\varepsilon_{k_z}-\varepsilon_{k'_z})
\, ,
\ee
where we defined $k'_z=k_z+q_z$.  
The expression in the curly brackets in Eq.~\eqref{Pi_2D_m}
agrees with the four-dimensional analogue of Eq.~(4) in
Ref.~\cite{CaronHuot:2007gq}.  The $\d$-function in \eqref{Pi_2D_m}
can be worked out explicitly as
\begin{equation}
\label{delta_2d}
\delta(\varepsilon_{k_z}-\varepsilon_{k_z'})
= \frac{\varepsilon_{q_z/2}}{|q_z|}\,\delta( k_z+q_{z}/2)\, ,
\end{equation}
which indicates that only a backward scattering $k_{z}=-k'_{z}=q_z/2$ 
contributes to the momentum diffusion of heavy quarks because of the
static limit in (1+1) dimensions.  We should note again that even such
backward scatterings were not allowed for massless quarks as already
discussed in Sec.~\ref{sec:massless}.

By performing the $k_z$ integration in Eq.~\eqref{Pi_2D_m} 
or by taking the $ \omega \to 0$ limit in Eq.~\eqref{eq:mass_th},
we find 
\be
\lim_{\omega\to0}\biggl[{\Im\Pi^{00}_{R,\LLL}(Q)\over\omega}
  \biggr] = \mq^2\, \frac{\pi s(\vq_{\perp})}{ T|q_z|\varepsilon_{q_z\over 2}}
    n_F(\varepsilon_{q_z\over 2})\bigl[
   1-n_F(\varepsilon_{q_z\over 2})\bigr] 
   \, ,
\label{reim}
\ee
where $ s({\bm q}_\perp)$ is defined in Eq.~\eqref{eq:dim_reduction}. 
Plugging this into Eq.~\eqref{GR_LLL} as before, we can obtain the
scattering rate $d\Gamma(\vq)/d^3\vq$, and then the finite mass
correction to $\kappa_{\parallel}$ as
\be
\label{kappa_m}
 \kappa_\parallel^{\rm LO,\,massive}
  = \frac{\alpha_s}{\pi}C^{\rm HQ}_{R}\mq^2\int d^3 \bm q\,\,
  \frac{s(\vq_{\perp})}{ \bigl[\bm q^2+s(\vq_{\perp})\bigr]^2}
  \,\frac{|q_{z}|}{\varepsilon_{q_z/2}}\,
  \frac{1}{1+\cosh(\varepsilon_{q_z/2}/T)}\,.
\ee
Now, it is clear from this expression that the dominant contribution
indeed comes from a region, $|\bm q| \sim (\alpha_s eB)^{1\over 2}$,
as claimed before.  In this region the Gaussian is approximated as
$e^{-\vq^2_{\perp}/(2eB)}\sim 1$, and we can replace $s(\vq_{\perp})$
by an effective Debye mass in the presence of the magnetic field as
\be
\label{mDB_def}
m_{D,B}^2\equiv s(\vq_{\perp}=0)
= 4\alpha_{s} T_{R} \sum_f \(\frac{|q_f eB|}{2\pi}\) 
\, . 
\ee
Furthermore, at the leading order in $\mq^2/\alpha_{s}eB$, 
we can approximate the quasi-energy as
\be
 \varepsilon_{q_z\over 2} = \sqrt{(q_z/2)^2+\mq^2} \sim |q_z/2|
\ee
in Eq.~\eqref{kappa_m}.  Since $q_z\sim (\alpha_s eB)^{1/2}\ll T$, we
can also make an approximation as
$\cosh(\varepsilon_{q_z/2}/T)\simeq 1$ in Eq.~\eqref{kappa_m}.
Putting those pieces together, we arrive at
 \begin{align}
\label{kappa_m1}
\kappa_\parallel^{\rm LO,\,massive} 
  &\simeq \frac{\alpha_s}{2\pi} C^{\rm HQ}_R \mq^2\int d^3 \bm q\,\,
  \frac{m_{D,B}^2}{ (\bm q^2+m_{D,B}^2)^2} \notag\\
  &= {\pi\over 2} \alpha_s C^{\rm HQ}_R \mq^2\, m_{D,B}
   = {1\over 2}\alpha_s  C^{\rm HQ}_R \mq^2 \sqrt{\alpha_s eB}
    \sqrt{2\pi T_R Q_{\rm em}}
\, . 
\end{align}
We should note that this result is independent of $T$ after dropping
terms in our assumed regime:  $\mq^2\ll \alpha_s eB\ll T^2\ll eB$.
Thus, if $\mq$ or $q_z\sim(\alpha_s eB)^{1/2}$ were comparable to $T$,
the mass correction would be a $T$ dependent function of $\mq/T$ and
$(\alpha_s eB)^{1/2}/T$, which are all dropped systematically in our
approximation.

\subsection{Thermal gluon contributions}
\label{sec:gluon}

We can capture the scatterings with thermal gluons by including the
imaginary part of the self-energy, $\Pi^{00}_{R,\gluon}$, coming from
hard thermal gluons.  A quick power counting shows that it is enough
to keep only the imaginary part of $\Pi^{00}_{R,\gluon}$, not the real
part, for a first non-vanishing contribution to $\kappa_\parallel$,
which we will refer to as ``leading order'' and will denote by
$\kappa_\parallel^{\rm LO,\,gluon}$.  The real part will be a
sub-leading correction to the leading-order screening mass from the quark loop 
$m_{D,B}^2\sim \alpha_s eB$, and we can neglect the gluon contribution in our regime. 
We will find that the final result of $\kappa_\parallel^{\rm LO,\,gluon}$ is
relatively suppressed by $T^2/(eB)$ compared to
$\kappa_\perp^{\rm LO}$ obtained in the preceding subsection.
Consequently, to leading order we
can replace Eq.~\eqref{eq:ReIm} with
\be
\label{eq:ImGR}
 \lim_{\omega\to0} (-2)\Im\biggl[{G_{R,{\gluon}}^{00}(Q)\over\omega}\biggr] =
  \lim_{\omega\to0}\,\frac{2}{\omega}\,
  {\Im\,\Pi_{R,\,{\rm gluon}}^{00}(Q) \over [Q^2-\Re\,\Pi_{R,\LLL}^{00}(Q)]^2}\, ,
\ee
neglecting the thermal gluon contribution to the screening mass as
compared to $\Pi_{R,\LLL}^{00}(Q)$ from the LLL quarks.

The dominant contribution to the imaginary part of the self-energy
comes from thermal gluons with hard momenta $\sim T$, which is
understood based on the interplay between phase space volume and the
Boltzmann suppression.  The dispersion relation of these hard gluons
could get modified in general by thermal effects.  Since
$\alpha_s eB\gg \alpha_s T^2$, however, the main source of the
correction appears from LLL quark loops, and in our regime,
$T^2\gg \alpha_s eB$, we should neglect such corrections and treat
hard gluons as free quasi-particles.

As a result, the imaginary part, $\Im\,\Pi_{R,\gluon}^{00}(Q)$, is
identical to the one without $B$, given by a cut of gluon one-loop
contribution to $\Pi^{00}_{R,\gluon}$ which is equal to the integrated
spectrum of scattering thermal gluons with a Coulomb vertex.
Equivalently, we can follow Ref.~\cite{Moore:2004tg} and work out
directly the $t$-channel scattering rate with thermal gluons with the
screened Coulomb propagator given above.  In this way the scattering
rate reads:
\be
\begin{split}
 &(2\pi)^3 2M_Q{d\Gamma_{\gluon}\over d^3\bm q} \\
 &= {1\over 2M_Q}\int {d^3 \bm k\over (2\pi)^3 2|\bm k|}
  {d^3\bm k'\over (2\pi)^3 2|\bm k'|} (2\pi)^4\delta^{(4)}(k'+Q-k)
  \big|{\cal M}\big|^2 n_B(|\bm k|)\bigl[1+n_B(|\bm k'|)\bigr]\,,
\end{split}
\label{scat}
\ee
where the $t$-channel amplitude with incoming and outgoing gluons of
color and polarization $(b,\epsilon^\mu)$ and $(c,\tilde\epsilon^\mu)$
is given by
\be
 {\cal M}^{bc} = -i4\pi\alpha_s f^{abc}
 \bigl[\bar U(P+Q)\gamma^0 t^a_R U(P)\bigr]
  G_{ra}^{00}(Q)(|\bm k|+|\bm k'|) (\epsilon\cdot \tilde\epsilon^*)\,,
\ee
where we included only $A^0$ Coulomb interaction for heavy quarks in
the static limit $P=(M_Q,\bm 0)$.  In this case the heavy quark
spinors can simplify as
\be
 \bar U(P+Q)\gamma^0 U(P)\simeq {\bar U}^\dag(P)U(P) = 2M_Q\,.
\ee
Color summation in the squared amplitude gives
\be
 \sum_{a,a',b,c} f^{abc}f^{a'bc} (t^a_R t^{a'}_R)=N_c \sum_a t^a_R t^a_R
 = N_c C_R^{\rm HQ} {\bf 1}\,,
\ee
and the polarization sum is
\be
 \sum_{\epsilon,\tilde\epsilon} |\epsilon\cdot \tilde\epsilon^*|^2
  = 1+\cos^2\theta_{\bm k \bm k'},
\ee
where $\theta_{\bm k\bm k'}$ is the angle between $\bm k$ and
$\bm k'$.  In the static limit we have $|\bm k'|=|\bm k|$ and  from this
the scattering rate becomes
\be
 {d\Gamma_{\gluon}\over d^3\bm q} = 4\alpha_s^2 N_c C_R^{\rm HQ}
  \int {d^3\bm k\over (2\pi)^3}\, \delta\bigl(|\bm k|-|\bm k-\bm q|\bigr)
  |G_{ra}^{00}(Q)|^2 (1+\cos^2\theta_{\bm k\bm k'})
   n_B(|\bm k|)\bigl[1+n_B(|\bm k|)\bigr]\,.
\ee
We can carry out the $\theta_{\bm k\bm k'}$-angle integration of $\bm k$ using
$\delta(|\bm k|-|\bm k-\bm q|)=|\bm q|^{-1}
\delta[\cos\theta_{\bm k\bm q}-|\bm q|/(2|\bm k|)] 
\Theta(|\bm k|-|\bm q|/2)$ and
$\cos\theta_{\bm k\bm k'}=1-|\bm q|^2/(2|\bm k|^2)$, and after all we
obtain
\be
 {d\Gamma_{\gluon}\over d^3\bm q}={\alpha_s^2\over\pi^2} N_c C_R^{\rm HQ}
  {|G_{ra}^{00}(Q)|^2\over |\bm q|}\int^{\infty}_{|\bm q|/2} dk\,
  k^2 \Biggl[ 1+\biggl(1-{\bm q^2\over 2k^2}\biggr)^2 \Biggr]
  n_B(k)\bigl[1+n_B(k)\bigr]\,,
\ee
where the screened Coulomb amplitude is
\be
 |G_{ra}^{00}(Q)|^2 = {1\over \bigl[\bm q^2+\Re\,\Pi^{00}_{R,\LLL}(Q)\bigr]^2}\,.
\ee
We note that in the computation of $\kappa_\parallel$ rotational
asymmetry arises only from
$s(\vq_\perp)=\Re\,\Pi^{00}_{R,\LLL}(\o=0,\vq)$ as defined in
Eq.~\eqref{eq:PiLLL}.

The Boltzmann suppression in ${d\Gamma/ d^3\bm q}$ restricts
$k\lesssim T$, and this in turn gives $|\bm q|\lesssim T$ from the
integration boundary.  Since $T^2\ll eB$, the asymmetric factor
$e^{-\bm q_\perp^2 / (2eB)}$ in $s(\vq_{\perp})$ is nearly the unity up to
corrections in powers of $T^2/(eB)$.  Therefore, at LO we recover
rotational symmetry that allows us to replace $(q_z)^2$ with
$q^2/3\equiv|\bm q|^2/3$, and we arrive at
\be
 \kappa^{\rm LO,\,gluon}_\parallel = {4\alpha_s^2\over 3\pi} N_c
  C_R^{\rm HQ} \int^\infty_0 dq\; {q^3 \over (q^2+m_{D,B}^2)^2}
  \int^\infty_{q/2} dk\; k^2\Biggl[ 1 + \biggl(1-{q^2\over 2k^2}
  \biggr)^2\Biggr] n_B(k)\bigl[1+n_B(k)\bigr]\,.
\ee
Apart from the value of the Debye mass $m_{D,B}^2 $
defined in Eq.~\eqref{mDB_def}, this integral is apparently identical to 
the conventional one without $B$ shown in Refs.~\cite{Moore:2004tg,CaronHuot:2007gq}.  
Therefore, the result of the integral can be obtained 
by simply substituting our $m_{D,B}$ for the conventional Debye mass in Refs.~\cite{Moore:2004tg,CaronHuot:2007gq} as 
\be
 \kappa^{\rm LO,\,gluon}_\parallel = {4\pi\alpha_s^2\over 9} N_c
  C_R^{\rm HQ} T^3 \biggl[ \log\biggl(\frac{1}{\alpha_s}\biggr)
  - \log\biggl( \frac{T_R Q_{\rm em} \,eB}{2\pi T^2}\biggr)
  +2\xi \biggr]\,,
\label{guon}
\ee
where $\xi=\frac{1}{2}-\gamma_E+\frac{\zeta'(2)}{\zeta(2)}\simeq -0.64718$.
This result is $T^2/(eB)$ smaller than $\kappa_\perp^{\rm LO}$ in Eq.~\eqref{LO}.

The above evaluation is systematic and consistent with our basic assumption $ \alpha_s eB \ll T^2 \ll eB $. 
First, as discussed in the beginning of this section, 
we neglected the thermal gluon contribution to the Debye mass $\sim gT \ll  m_{D,B}$. 
Next, in Refs.~\cite{Moore:2004tg,CaronHuot:2007gq}, 
the authors obtained the LO result from the contributions of the hard thermal gluons $ \gtrsim T$, 
and neglected corrections of the order of $ m_D^2/T^2 \sim g^2 $ 
from the contributions of the soft gluons $ \sim m_D$. 
In our case, we can also neglect these corrections 
$ \sim m_{D,B}^2/T^2 \sim \alpha_s eB/T^2 \ll1$ along with the above hierarchy 
in the present analysis at the LO accuracy. 
We leave studies of the higher-order contributions for future work, which also have 
relevance to the QCD Kondo effect recently discussed in Refs.~\cite{Hattori:2015hka,Ozaki:2015sya}.


It is instructive to compare the LO hard thermal gluon contribution
\eqref{guon} with the LO massive light quark contribution to
$\kappa_\parallel$ in Eq.~\eqref{kappa_m1}.  The ratio is found to be
\be
 {\kappa_\parallel^{\rm LO,\,massive} \over
  \kappa_\parallel^{\rm LO,\,gluon}} \sim
 {\alpha_s (\alpha_s eB)^{1/2} \mq^2 \over \alpha_s^2 T^3}
 = \biggl({\mq^2 \over \alpha_s eB}\biggr)
 \biggl({\alpha_s eB\over T^2}\biggr)^{1/2}
 \biggl({eB\over T^2}\biggr)\,.
\ee
The first two factors are small according to our working regime, but
the last factor can be large.  Therefore, the massive contribution
$\kappa_\parallel^{\rm LO,\, massive}$ could be in principle as
comparably large as $\kappa_\parallel^{\rm LO,\, gluon}$, and this
happens when $eB\sim \alpha_s(T^6/\mq^4)$.  Then, to be consistent
with our assumed regime, $\alpha_s\,eB\ll T^2$, we have a constraint
of $\alpha_s\ll \mq^2/T^2$, which is not quite likely true in the heavy ion
collisions.  Hence, in realistic heavy ion experiments,
$\kappa_\parallel^{\rm LO,\,gluon}$ is a dominant contribution to the longitudinal
diffusion coefficient.

\section{Phenomenological implications}
\label{sec:FP}

In the previous sections, we have computed the heavy quark momentum diffusion
coefficients, $\kappa_\perp$ and $\kappa_\parallel$, in the QGP in the
presence of strong magnetic field $eB\gg T^2$ at LO in $\alpha_s$, 
and have found
\begin{equation}
\label{kappa_ratio}
 \frac{\eta_\parallel(B)}{\eta_\perp(B)}
  = \frac{\kappa_\parallel(B)}{\kappa_\perp(B)} \sim
  \frac{T^2}{eB} \ll 1\, .
\end{equation}
We now study the phenomenological implications of
Eq.~\eqref{kappa_ratio}.

\begin{figure}
     \begin{center}
              \includegraphics[width=0.3\hsize]{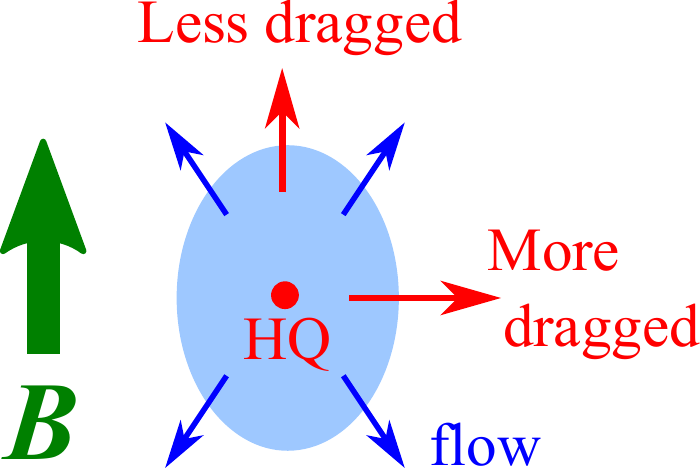}
     \end{center}
\vspace{-0.5cm}
\caption{Schematic illustration for the mechanism to derive an
              additional contribution to the elliptic flow of heavy
              quarks.  Because of $\kappa_\perp \gg \kappa_\parallel$
              heavy quarks are more dragged along the in-plane than
              along the out-of-plane.}
\label{fig:schematic1}
\end{figure}

To give a (semi-)quantitative estimate of its influence on the
elliptic flow of heavy quarks, we will implement the anisotropic
$\kappa_{\perp,\parallel}$ in description of the evolution of an open heavy
quark in the expanding QGP (see Fig.~\ref{fig:schematic1} for a
schematic illustration of our physical picture).  Following
conventions in the heavy ion collision literature, we will take the
in-plane and out-of-plane direction as $x$- and $y$-direction,
respectively.  We will assume an external magnetic field along the
$y$-direction.  Therefore, $\kappa_{xx}=\kappa_x=\kappa_\perp$ and
$\kappa_{yy}=\kappa_y=\kappa_\parallel$.  In realistic situations in the heavy ion
collisions, the background flow $u_{x,y}$ of plasma fireball depends
on space and time.  In what follows we limit ourselves to some spatial
regions where we can treat $u_{x,y}$ as spatially homogeneous fields.

The evolution of a heavy quark is described by the Langevin equation
 with the homogeneous flow effects ($M_Q$ is the heavy quark mass)
 \be
 \label{Langevin_B}
  \frac{d p_x}{d\tau} = -\eta_{x}(\tau)
 \Big[ p_x- M_Q u_{x}(\tau)\Big]+ \xi_x(\tau)\,,
 \qquad
 \frac{d p_y}{d\tau} = -\eta_{y}(\tau)
 \Big[ p_y- M_Q u_{y}(\tau)\Big]+ \xi_y(\tau)
\ee
with
\be
\label{noise_B}
 \langle \xi_{x}(\tau) \xi_x(\tau')\rangle=\kappa_{x} \delta(\tau-\tau')\,,
 \qquad
 \langle \xi_{y}(\tau)\xi_{y}(\tau')\rangle=\kappa_{y}
  \delta(\tau-\tau')\,.
\ee
Equivalent to Eq.~\eqref{Langevin_B},  we can translate these equations into the Fokker-Planck equation as
\be
\label{FP_B}
\begin{split}
 \pd_{\tau} P(p_x, p_y;\tau)
 &= - \Bigl[ \eta_{x}(\tau)\pd_{p_{x}}\bigl\{\[ p_{x}-M_Q
 u_{x}(\tau)\]+\[M_QT(\tau)\]\pd_{p_{x}}\bigr\} \\
&\qquad + \eta_{y}(\tau)\pd_{p_{y}}\bigl\{
  \(p_{y}-M_Q u_{y}(\tau)\)+\(M_QT(\tau)\)\pd_{p_{y}}\bigr\}
   \Bigr] P(p_{x},p_{y};\tau)\, , 
\end{split}
\ee
where $P(p_x, p_{y};\tau)$ denotes the probability of finding a heavy
quark at $p_x$ and $p_y$, and $T(\tau)$ is the time-dependent temperature of the background plasma.

The Green's function to Eq.~\eqref{FP_B}, i.e.\ the probability of
finding a heavy quark in $(p_{x},p_{y})$ at time $\tau$ under the initial condition in
$(p^0_x,p^0_y)$ can be found analytically as
\be
\label{FP_Green}
 \<p_{x},p_{y}|p^{0}_{x},p^{0}_{y}\> = \prod_{i=x,y}
\frac{1}{\sqrt{2\pi \Delta_{x}(\tau)}}\exp
 \Biggl\{-\frac{\[p_i - \bar{p}_i(\tau)-p^0_i\,e^{-\Gamma_i(\tau)}\]^2}
  {2\Delta_i(\tau)}\Biggr\} \,.
\ee
Here we introduced new variables:
\begin{align}
 \label{FP_parameter}
  \Gamma_i(\tau) &\equiv \int^{\tau}_{\tau_0} d\tau'\, \eta_i(\tau')\, ,\\
  \bar{p}_i(\tau) &\equiv M_Q\, e^{-\Gamma_{i}(\tau)}\int^{\tau}_{\tau_0} d\tau'\,
   e^{\Gamma_i(\tau')}\, \eta_{i}(\tau') u_{i}(\tau') \, , \\
 \Delta_{i}(\tau) &\equiv 2 M_Q\, e^{-2\Gamma_{i}(\tau)}
  \int^{\tau}_{\tau_0} d\tau'\, e^{2\Gamma_{i}(\tau')}\,
  \bigl[T(\tau')\eta_i(\tau')\bigr]\, . 
\end{align}
With Eq.~\eqref{FP_Green}  the solution to Eq.~\eqref{FP_B} under the
initial condition $P_0(p_{x}, p_{y};\tau_0)$ can be written as
\begin{equation}
 \label{FP_sol}
 P(p_{x},p_{y};\tau)
 = \int dp^{0}_{x}\,dp^{0}_{y}\, \<p_x,p_y|p^0_{x},p^0_{y}\>\,
  P_0(p_x^0, p_y^0;\tau_0)\, . 
\end{equation}

The physical meaning of each term in Eq.~\eqref{FP_parameter} is
rather transparent:  $\Gamma_i(\tau)$ is the effective damping
factor which will wash out the memory of earlier distribution of heavy
quarks.  Indeed, a large value of $\Gamma_i(\tau)$ would suppress $p^0_i$
dependence in the Green's function.  $\bar{p}_i$ is nothing but the
solution to the Langevin equation~\eqref{FP_B} with homogeneous initial
condition $p_i=0$ after averaging over the noise.  It characterizes
the heavy quark flow due to dragging by the expanding QGP medium.
Finally, $\Delta_i(\tau)$ is generated by the noise during the
Langevin dynamics, which would blur the information contained in the
initial distribution.
 
\begin{figure}
     \begin{center}
              \includegraphics[width=0.6\hsize]{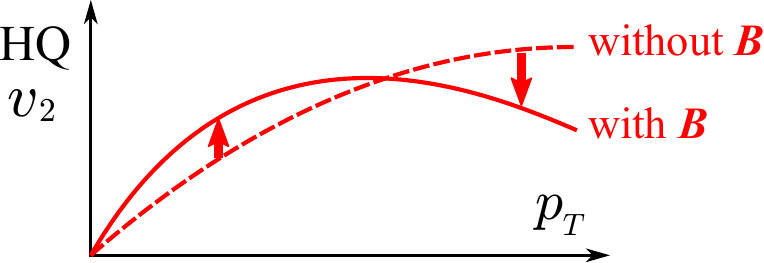}
     \end{center}
\vspace{-0.5cm}
\caption{Schematic illustration for the effects of the magnetic field
              on the heavy quark elliptic flow.}
\label{fig:schematic2}
\end{figure}

The modified distribution of heavy quarks has a characteristic
structure as illustrated in Fig.~\ref{fig:schematic2}.  For low
momentum, i.e.\ $p_T\lesssim |\boldsymbol{u}|M_Q$, the anisotropy in
$\eta_{\perp,\parallel}$ or $\kappa_{\perp,\parallel}$ gives rise to a
positive contribution to $v_2$ of heavy quarks.  This is because, when
$\eta_x\gg \eta_y$, heavy quarks will gain more momenta in the
$x$-direction than that in the $y$-direction, which is embodied in
$\bar{p}_i$ terms in Eq.~\eqref{FP_Green}.  In contrast, for high
momentum, i.e.\ $p_T\gtrsim |\boldsymbol{u}|M_Q$, the magnetically 
induced contribution to the $v_2$ is opposite.  This is because, for an isotropic
initial distribution, more memory is washed out in the $x$-direction
than in the $y$-direction, which is embodied in $p_i\,e^{-\Gamma_i}$
terms in Eq.~\eqref{FP_Green}.  More elaborate numerical simulations are in progress.

\section{Summary}
\label{sec:discussions}

In this work, 
 we have computed heavy quark momentum diffusion rate $\kappa$
 of a quark-gluon plasma in the presence of strong magnetic fields $eB\gg T^2$ 
at the leading order in $\alpha_{s}$. 
While the contribution from thermal gluons is still isotropic at 
the leading order in $T^{2}/eB$ (cf.~section~\ref{sec:gluon}), 
we found that the fermionic contribution becomes anisotropic.
Indeed, in the massless limit, the fermionic contribution to 
the longitudinal diffusion $\kappa_{\parallel}$ vanishes 
under the LLL approximation (cf.~section~\ref{sec:massless}), 
while their contributions in the transverse direction 
shown in Eq.~\eqref{LO} is non-vanishing 
$\kappa_{\perp}\sim \alpha_{s}^2 eB T$ 
and is dominant over the gluonic contributions.
As a result, we have a large anisotropy 
\begin{equation}
\label{kappa_ratio2}
\frac{\etapa}{\etape}=\frac{\kapa}{\kape}\sim \frac{T^{2}}{eB}\ll1\, . 
\end{equation}
We call this anisotropy in the drag force coefficients ``magnetic drag anisotropy.''

Turning to the phenomenological implications of ``magnetic drag anisotropy'', 
we first recall that for heavy quarks in an expanding plasma, 
the drag force will push them to co-move with the medium.    
For low momentum heavy quarks, the anisotropy 
$\eta_{\parallel}<\eta_{\perp}$ implies that those heavy quarks will gain more momentum in the in-plane direction than in the out-of-plane direction, 
as the magnetic field points to the out-of-plane direction.
Therefore ``magnetic drag anisotropy'' will generate {\it positive} elliptic flow $v_{2}$ 
for those low $p_{T}$ heavy quarks. 

A body of conventional study on heavy quark dynamics is based on isotropic drag coefficients. 
In those studies~\cite{HFR,Bera,Sca}, there are some tensions in 
simultaneously describing the nuclear modification factor $R_{AA}$ 
and the elliptic flow $v_{2}$ of open heavy flavors in the low $p_{T}$ regime. 
If one tries to reproduce the experimentally measured $R_{AA}$ 
which is not significantly suppressed in this regime (see Ref.~\cite{review_heavy} for a review), 
the estimate of the resulting $v_{2}$ 
typically undershoots the experimental data. 
This is because, as pointed out in Ref.~\cite{Moore:2004tg}, 
 $R_{AA}$ and $v_{2}$ are tightly correlated; namely, 
when suppression of $R_{AA}$ is moderate, 
the thermalization of heavy quarks takes a long time, 
meaning a significantly small $v_2$ of heavy quarks 
compared to that of the medium. 
A common assumption involved in such estimates is the isotropy of the drag coefficients. 
It is thus tempting to propose a new scenario for resolution of this issue, 
the so-called ``heavy-flavor puzzle'', on the basis of 
the anisotropic drag coefficients $\eta_{\parallel}\ll \eta_{\perp}$ shown in the present work. 
As discussed in Sec.~\ref{sec:FP}, 
the anisotropic drag force coefficient will be able to generate an additional positive 
contribution to the elliptic flow in the low $p_{T}$ regime without significantly changing $R_{AA}$. 
%
%
%
While our discussion on the consequence of anisotropic drag force coefficients would apply 
to any microscopic mechanism which would induce $\eta_\parallel \ll \eta_\perp $, 
we indeed identified one such origin of the mechanism, namely the strong magnetic field. 
Quantitative study on the basis of the dynamical modeling discussed in Sec.~\ref{sec:FP} 
will be interesting future work. 
Our results for $\eta$ can readily be implemented in those computations.


As the last comment, magnetic drag anisotropy discussed here 
has a deep connection to the non-dissipative nature of anomalous transport~\cite{RS,SYee,limitingv} 
and this connection deserves a further study.
We leave those interesting directions for the future study. 

\section*{Acknowledgments}
We thank Yukinao~Akamatsu, Koichi~Murase, Jorge Noronha, Hiroshi Ohno, Rob Pisarski, Alexander Rothkopf, Bjoern Schenke, and Sayantan Sharma for helpful discussions.
K.~F.\ is supported by JSPS KAKENHI Grant No.\ 15H03652 and 15K13479, K.H. is supported by JSPS Grants-in-Aid No.~25287066, and 
Y.~Y. is supported by DOE Contract No.~DE-SC0012704.


\end{document}